\input harvmac
\def\half{{1 \over 2}}
\def\dzm{{\partial}}

\def\p{{\partial}}
\def\s{{\sigma}}

\def\L{{\Lambda}}

\def\a {{\alpha}}
\def\b {{\beta}}

\def\g {{\gamma}}
\def\d {{\delta}}
\def\e {{\epsilon}}

\def\ad {{\dot\alpha}}
\def\bd {{\dot\beta}}

\def\t {{\theta}}
\def\ta {{\theta^\alpha}}

\def \ad {{\dot \a}}
\def \bd {{\dot \b}}
\def \t {{\theta}}
\def \tb {{\overline\theta}}

\Title{\vbox{\hbox{ IFT-P.067/2001}
\hbox{ CITUSC/01-034}
\hbox{ CALT-68-2354}}}
{\vbox{\centerline{One-Loop N-Point Superstring Amplitudes}
\smallskip
\centerline{with Manifest d=4 Supersymmetry}}}
\centerline{Nathan Berkovits,$^{\spadesuit}$\foot{E-Mail: nberkovi@ift.unesp.br}
and Brenno Carlini Vallilo$^{\spadesuit\diamondsuit}$\foot{E-Mail: 
vallilo@ift.unesp.br, vallilo@theory.caltech.edu}}
\bigskip
\bigskip
\centerline{\it $^{\spadesuit}$   
Instituto de F\'\i sica Te\'orica, Universidade Estadual
Paulista}
\centerline{\it Rua Pamplona 145, 01405-900, S\~ao Paulo, SP, Brasil}
\smallskip
\centerline{\it $^{\diamondsuit}$
California Institute of Technology, Pasadena, CA 91125, USA}
\vskip .2in
The hybrid formalism for the superstring is used to compute 
one-loop amplitudes with an arbitrary number of external 
d=4 supergravity states. These 
one-loop N-point amplitudes are expressed as Koba-Nielsen-like
formulas with manifest d=4 supersymmetry.

\Date{October 2001}
\newsec {Introduction}

Although some nonperturbative aspects of superstring theory can now be
studied through
duality symmetries, there are still many perturbative
aspects of superstring theory
which are not well understood. For example, perturbative
finiteness of superstring amplitudes has only been proven
by patching
together amplitude computations using the light-cone Green-Schwarz (GS)
formalism with amplitude computations using the Ramond-Neveu-Schwarz (RNS)
formalism \ref\Man{S. Mandelstam, ``The N Loop String Amplitude:
Explicit Formulas, Finiteness and Absence of Ambiguities'',
Phys. Lett. B277 (1992) 82.}.\foot
{The proof of \ref\twistor{N. Berkovits, ``Finiteness 
and Unitarity of Lorentz-Covariant
Green-Schwarz Superstring Amplitudes", Nucl. Phys. B408 (1993) 43,
hep-th/9303122.} 
using the twistor-string formalism is incorrect since
the unphysical poles come from the chiral boson correlation function
were not treated in a BRST-invariant manner.}
This patching is necessary
since the light-cone GS amplitudes have contact term problems whereas
the RNS amplitudes are only spacetime supersymmetric after summing
over spin structures. 

Another aspect of perturbative superstring theory which is not well
developed is the computation of scattering amplitudes involving
more than four 
external fermions. After extracting their low-energy contributions,
these amplitude computations could be useful for simplifying 
analogous computations in QCD \ref\Bern{Z. Bern and
D. Kosower, ``The Computation of Loop Amplitudes in Gauge Theories'',
Nucl. Phys. B379 (1992) 452.}. However, 
contact term interactions in the light-cone GS formalism and the
complicated nature of Ramond vertex operators in the RNS formalism have
made these amplitudes difficult to compute.
So to study perturbative finiteness and to obtain
amplitude expressions for more than four external fermions, it
would be nice to have a superstring formalism which does not
suffer from the problems of the GS and RNS formalisms.

Over the last seven years, a ``hybrid'' formalism has been developed
which combines the advantages of the GS and RNS formalisms without
including their disadvantages.
The hybrid formalism
is manifestly spacetime supersymmetric
and therefore does not require summing over spin structures
and can easily handle an arbitrary number of external fermions.
Furthermore, in a flat target-space background, 
the hybrid worldsheet action is quadratic so scattering amplitudes
can be computed using free field OPE's.

In this paper, one-loop scattering amplitudes will be computed
using the d=4 version of the hybrid formalism which describes the
Type II superstring\foot{Although only the Type II superstring
will be discussed here, all results are easily generalized
to one-loop open or heterotic superstring amplitudes.}
compactified on any six-dimensional
manifold which preserves at least d=4 supersymmetry
\ref\hybfour{N. Berkovits, ``Covariant Quantization of the Green-Schwarz
Superstring in a Calabi-Yau Background'', Nucl. Phys. B431 (1994) 258,
hep-th/9404162\semi
N. Berkovits, ``A New Description of the Superstring'', 
proceedings of VIII J. A. Swieca Escola de Vera\~o,
hep-th/9604123.}.
These one-loop amplitudes will be computed for an arbitrary number
of external N=2 d=4 supergravity states
which are independent of the compactification. The amplitudes will
be expressed as Koba-Nielsen-like formulas with manifest N=2
d=4 supersymmetry which closely resemble the formulas computed in
\ref\kn{N. Berkovits, ``Super-Poincar\'e Invariant Koba-Nielsen
Formulas for the Superstring'', Phys. Lett. B385 (1996) 109, hep-th/9604120.}
for tree amplitudes.
Although these one-loop amplitudes could be computed in principle using
the RNS or light-cone GS formalisms, the problems described earlier have
up to now prevented such computations. However, it will be shown that
when all external states are NS-NS, the amplitudes agree with
the RNS result. 

There are various generalizations of these one-loop computations which
might be possible. One possible generalization would be to compute
one-loop amplitudes involving external compactification-dependent
states. Such computations would be interesting for anomaly analysis
but require a better understanding of correlation
functions involving the chiral boson $\rho$ in the hybrid formalism.
A second possible generalization would be to compute one-loop
amplitudes using the d=6 
\ref\hybsix
{N. Berkovits, ``Quantization of the Type II Superstring in
a Curved Six-Dimensional Background'', Nucl. Phys. B565 (2000) 333,
hep-th/9908041.} or d=10 
\ref\hybpure{N. Berkovits, 
``Super-Poincar\'e Covariant Quantization of the
Superstring'', JHEP 0004 (2000) 018, hep-th/0001035.}
pure spinor versions of the hybrid
formalism which manifestly preserve more spacetime symmetries.
And a third possible generalization would be to use the hybrid
formalism to compute multiloop amplitudes which might be useful
for studying perturbative finiteness. Up to now, only special 
``topological'' amplitudes
\ref\topamp{
M. Bershadsky, S. Cecotti, H. Ooguri and C. Vafa, ``Holomorphic
Anomalies in Topological Field Theories'', Nucl. Phys. B405 (1993) 279,
hep-th/9302103\semi
I. Antoniadis, E. Gava, K.S. Narain
and T.R. Taylor, ``Topological Amplitudes in String Theory'',
Nucl. Phys. B413 (1994) 162, hep-th/9307158\semi
M. Bershadsky, S. Cecotti, H. Ooguri and C. Vafa, ``Kodaira-Spencer
Theory of Gravity and Exact Results for Quantum String Amplitudes'',
Comm. Math. Phys. 165 (1994) 311, hep-th/9309140.}
which involve trivial correlation functions
of the chiral boson $\rho$ have been computed at higher loops using
the hybrid formalism \ref\topo{N. Berkovits and C. Vafa,
``N=4 Topological Strings'', Nucl. Phys. B433 (1995) 123, hep-th/9407190.}.

Section 2 of this paper will review the d=4 hybrid formalism.
Section 3 will use the correlation functions of the
hybrid worldsheet variables on a torus to compute explicit 
d=4 supersymmetric expressions 
for one-loop N-point amplitudes with external massless d=4 states. 
And section 4 will prove that these one-loop expressions are
gauge-invariant, single-valued, modular invariant,
and agree
with the RNS one-loop amplitudes when all external states are 
in the NS-NS sector.

\newsec{Review of d=4 Hybrid Formalism}

\subsec{Worldsheet action}

After embedding the RNS superstring in a $\hat c=2$ N=2 superstring and
performing a field redefinition on the worldsheet variables,
the superstring can be described in a d=4 super-Poincar\'e covariant
manner using the d=4 hybrid formalism \hybfour. This formalism can be used 
to describe either the uncompactified superstring or any compactification
of the superstring which
preserves at least d=4 supersymmetry. For the Type II superstring,
the worldsheet variables in this formalism consist of the N=2 d=4
superspace variables $[x^m,\t^\a_L,\tb^\ad_L,\t^\a_R,\tb^\ad_R]$ for
$m=0$ to 3 and $(\a,\ad)=1$ to 2, the fermionic conjugate momenta
$[p_{L\a},\bar p_{L\ad}, 
p_{R\a},\bar p_{R\ad}]$, a left and right-moving chiral boson
$[\rho_L,\rho_R]$, and a worldsheet N=(2,2) $\hat c=3$ superconformal field
theory which represents the compactification manifold.

The worldsheet action in conformal gauge for these fields is
\eqn\act{\int dz d\bar z(\half\p x^m\bar\p x_m +p_{L\a}\bar\p\t_L^\a
+\bar p_{L\ad}\bar\p\tb_L^\ad 
+p_{R\a}\p\t_R^\a +
\bar p_{R\ad}\p\tb_R^\ad)}
$$ + S_{\rho_L} + S_{\rho_R}
+S_C$$
where $S_{\rho_L}$ and $S_{\rho_R}$ are the actions for the
left and right-moving chiral bosons,
and $S_C$ is the action for the compactification-dependent variables.
Note that $z$ versus $\bar z$ is correlated with $L$ versus
$R$, and not with $\t$ versus $\tb$.

As $y\to z$,
the free-field OPE's of the four-dimensional variables are 
\eqn\ope{x^m(y) x^n(z)\to - \eta^{mn}\log|y-z|^2,}
$$\rho_L(y) \rho_L(z) \to -\log(y -z),
\quad \rho_R(y) \rho_R(z) \to -\log(\bar y -\bar z),$$
$$p_{L\a}(y)\theta_L^\b (z)\to {\delta_\a^\b\over{y -z}},\quad
\bar p_{L\ad}(y)\bar\theta_L^\bd (z)\to {\delta_\ad^\bd\over{y -z}},$$
$$p_{R\a}(y)\theta_R^\b (z)\to {\delta_\a^\b\over{\bar y -\bar z}},\quad
\bar p_{R\ad}(y)\bar\theta_R^\bd (z)\to 
{\delta_\ad^\bd\over{\bar y -\bar z}}.$$
Note that all worldsheet variables are periodic
and that the chiral boson $\rho$ can not
be fermionized since
$e^{\rho_L(y)}~e^{-\rho_L(z)}~\to 
(y -z)$ and
$e^{\rho_R(\bar y)}~e^{-\rho_R(\bar z)}~\to 
(\bar y -\bar z)$. It has the same behavior as the
negative-energy field $\phi$ that appears when bosonizing the RNS ghosts.

\subsec{Worldsheet N=2 superconformal generators}

Using RNS variables, one can construct
the twisted $\hat c=2$ N=2 superconformal generators
$[T,G,\bar G,J]=[T_{matter}+T_{ghost}, j_{BRST}, b, b c 
+ \xi\eta]$
where the $L/R$ index is being suppressed in this subsection.
In terms of the d=4 hybrid variables, these
generators are mapped under the field redefinition
to\hybfour
\eqn\gen{T=-\half\p x^m \p x_m -
p_{\a}\p \t^\a - \bar p_{\ad }\p\tb^\ad -\half(\p\rho\p\rho
+\partial^2 \rho) +T^C,}
$$G=e^{\rho} (d)^2 +G^C, \quad
\bar G=e^{-\rho} (\bar d)^2 +\bar G^C, \quad
J=-\dzm\rho +J^C,
$$
where
$$d_{\a}=p_{\a}+{i\over 2}\tb^\ad\dzm x_{\a\ad}-
{1\over 4}(\tb)^2\dzm\t_{\a}
+{1\over {8}}\t_{\a} \dzm (\tb)^2,$$
$$\bar d_{\ad}=\bar p_{\ad}
+{i\over 2}\t^\a\p x_{\a\ad}-{1\over 4}(\t)^2\p\tb_{\ad}
+{1\over {8}}\tb_{\ad} \dzm (\t)^2,$$
$x_{\a\ad}=x_m \sigma_{\a\ad}^m$,
and $[T^C,G^C,\bar G^C,J^C]$ 
are the twisted $\hat c=3$ N=2 generators of
the superconformal field theory
used to describe the compactification manifold.

As was shown by Siegel\ref\sieg{W. Siegel, ``Classical Superstring
Mechanics'', Nucl. Phys. B263 (1986) 93.},
$d_{\a}$ and $\bar d_{\ad}$ satisfy
the OPE's
\eqn\dope{d_{\a} (y) \bar d_{\ad}(z) \to i{{\Pi_{\a\ad}}\over{y -z}}, \quad
d_{\a}(y) d_{\b}(z) \to {\rm regular},\quad 
\bar d_{\ad}(y) \bar d_{\bd}(z) \to {\rm regular},}
$$ d_\a(y) U(z) \to
{{D_\a U}\over{y-z}},\quad 
\bar d_\ad(y) U(z) \to
{{\bar D_\ad U}\over{y-z}},$$
$$ d_\a(y) \p\t^\b(z) \to {{\d_\a^\b}\over{(y-z)^2}},
\quad
\bar d_\ad(y) \p\tb^\bd(z) \to {{\d_\ad^\bd}\over{(y-z)^2}},
$$
$$
d_{\a}(y) \Pi^m(z) \to -i{{\sigma^m_{\a\ad} \dzm\tb^\ad}\over{y -z}},\quad
\bar d_{\ad}(y) 
\Pi^m(z) \to -i{{\sigma^m_{\a\ad} \dzm\t^\a}\over{y -z}},\quad
\Pi^m(z)\Pi^n(z)\to -{{\eta^{mn}}\over{(y-z)^2}}$$
where
\eqn\pidefn{\Pi^m=
\dzm x^m -{i\over 2}\s^m_{\a\ad}(\t^\a \dzm\tb^\ad+
\tb^\ad\dzm\t^\a),\quad
D_{\a}=\p_{\t^\a}-{i\over 2}\tb^\ad\p_{\a\ad},\quad
\bar D_{\ad}=\p_{\tb^\ad}-{i\over 2}\t^\a\p_{\a\ad},}
and $U(z)=U(x(z),\t(z),\tb(z))$ is a scalar superfield.
The advantage of working with the variables $d_{\a}$,
$\bar d_{\ad}$ and $\Pi^m$ is that they
commute with the spacetime supersymmetry generators
$$q_{\a}=\int dz [p_{\a} -{i\over 2}
\tb^\ad\dzm x_{\a\ad}-{1\over {8}}(\tb)^2\dzm\t_{\a}],
\quad \bar q_{\ad}=\int dz [\bar p_{\ad}
-{i\over 2}
\ta\dzm x_{\a\ad}-{1\over {8}}(\t)^2\dzm\tb_{\ad}].
$$

\subsec{Massless compactification-independent vertex operators}

Since the hybrid formalism is a critical N=2 superconformal field
theory, physical vertex operators can be described by U(1)-neutral
N=2 primary fields with respect to the superconformal generators of \gen.
For massless states of the Type II superstring which are independent
of the compactification, the unintegrated vertex operator only
depends on the zero modes of $[x^m,\t_L^\a,\tb_L^\ad,\t_R^\a,\tb_R^\ad]$
and is therefore described by the scalar superfield
$U(x,\t_L,\tb_L,
\t_R,\tb_R)$, which is the prepotential for an N=2 d=4 supergravity
and tensor multiplet
\ref\eff{N. Berkovits and W. Siegel,
``Superspace Effective Actions for 4D Compactifications
of Heterotic and Type II Superstrings'',
Nucl. Phys. B462 (1996) 213,
hep-th/9501016.}. 
The NS-NS fields for the graviton, anti-symmetric
tensor, and dilaton are in the $\t_L^\a\tb_L^\ad\t_R^\b\tb_L^\bd$ 
component of $U$,
i.e.
\eqn\nsns{
[D_{L\a},\bar D_{L\ad}]
[D_{R\b},\bar D_{R\bd}] U|_{\t=\tb=0} =
\s^m_{\a\ad}\s^n_{\b\bd}
(h_{mn}+b_{mn}+\eta_{mn}\phi),}
and the R-R field strengths
for the U(1) vector 
and complex scalar are in the 
$(\t_L)^2\tb_L^\ad$
$ (\t_R)^2\tb_R^\bd$
and $\t_L^\a(\tb_L)^2 $
$(\t_R)^2\tb_R^\bd$ components of $U$, i.e.
$$(D_L)^2 \bar D_{L\ad}(D_R)^2\bar D_{R\bd}U|_{\t=\tb=0} =
(\s^{mn})_{\ad\bd} F_{mn}, \quad
D_{L\a}(\bar D_L)^2(D_R)^2\bar D_{R\bd}U|_{\t=\tb=0} =\s^m_{\a\bd}\p_m y,$$
where $D_{L\a}$ and 
$\bar D_{L\ad}$ are defined as in \pidefn\ with $\t_L^\a$ and $\tb_L^\ad$ 
variables,
and $D_{R\a}$ and 
$\bar D_{R\ad}$ are defined as in \pidefn\ with $\t_R^\a$ and $\tb_R^\ad$ 
variables.
Note that
in standard SU(2) notation for N=2 superspace,
$\t_L^\a=\t_+^\a$, 
$\tb_L^\ad=\tb^{+\ad}$, 
$\t_R^\a=\t_-^\a$, and
$\tb_R^\ad=\tb^{-\ad}$. 

For $U$ to be an N=2 primary field, it must satisfy the constraints
$$(D_L)^2 U=(\bar D_L)^2 U=(D_R)^2 U=(\bar D_R)^2 U=\p_m \p^m U=0.$$
The first four constraints are the N=2 
d=4 supersymmetric generalization of the
usual polarization conditions, and the last constraint is the equation of
motion in this gauge.
Using the OPE's of \ope, one can compute that
the integrated form of the closed superstring vertex operator is 
\eqn\closedv{\int d^2 z~ V = \int d^2 z ~
\bar G_L (G_L (\bar G_R (G_R(U))))     
=\int d^2 z | H(z) |^2 U(z,\bar z)}
where 
\eqn\defHH{H(z)=
d_L^\a(z) (\bar D_L)^2 D_{L\a}  +
\bar d_L^\ad(z) (D_L)^2\bar D_{L\ad}  }
$$+
\p\t_L^\a(z) D_{L\a}
-\dzm\tb_L^\ad(z) \bar D_{L\ad}+{i\over 2} \Pi_{L\a\ad}(z)[D_L^\a,
\bar D_L^\ad],$$
$|~~|^2$ signifies the left-right product, and
all superspace derivatives $D_\a$ and $\bar D_\ad$
act on the superfield $U$.

Using the field
redefinition to write \closedv\ in terms of RNS variables, one can check
that the NS-NS fields of \nsns\ couple in \closedv\ as
\eqn\vertrns{\int d^2 z ~V=\int d^2 z 
(\p x^m +i\psi_L^m\psi^p_L k_p)
(\bar\p x^n +i\psi_R^n\psi^q_R k_q)
(h_{mn}+b_{mn}+\eta_{mn}\phi),}
which is the usual RNS vertex operator.
Furthermore, $\int d^2 z ~V$ is invariant up to a surface term under
the linearized 
gauge transformation 
\eqn\ling{\d U = (D_L)^2\L_L +(\bar D_L)^2 \bar\L_L
+ (D_R)^2\L_R +(\bar D_R)^2 \bar\L_R,}
which is the N=2 d=4 supersymmetric
generalization of the gauge transformation
$\d(h_{mn}+b_{mn}+\eta_{mn}\phi) = \p_m \lambda_{L n}
+ \p_n \lambda_{Rm}$.
For example, under $\d U= (D_L)^2\L_L$, 
\eqn\changeg{\d V = (-\p\tb_L^\ad \bar D_{L\ad}+{i\over 2}\Pi^m_{L\a\ad}
[D_L^\a,\bar D_L^\ad])\bar H(\bar z) (D_L)^2 \L_L}
$$= 
(-\p\t_L^\a D_{L\a} -\p\tb_L^\ad \bar D_{L\ad}-\Pi^m_L\p_m)\bar H(\bar z)
(D_L)^2\L_L
=-\p_z (\bar H(\bar z) (D_L)^2\L_L).$$

\newsec{One-Loop N-point Scattering Amplitude}

\subsec{Topological prescription}

In \topo, a ``topological'' prescription was given for computing
scattering amplitudes for any $\hat c=2$ N=2 superconformal field theory.
This prescription uses the twisted version of the N=2 superconformal field
theory in which the worldsheet N=2 superconformal ghosts contribute
zero central charge and decouple from the scattering amplitudes. 
For the $\hat c=2$ N=2 superconformal field theory representing the
self-dual string on $T^2\times R^2$, this prescription was used by
Ooguri and Vafa \ref\alll{H. Ooguri and C. Vafa,
``All Loop N=2 String Amplitudes'', Nucl. Phys. B451 (1995)
121, hep-th/9505183.}
to explicitly compute the $g$-loop partition
function dependence on the $T^2$ moduli.

For the $\hat c=2$ N=2 superconformal field theory which represents
the superstring using the
d=4 hybrid formalism reviewed in section 2, there is a subtlety
in the prescription caused by the negative energy chiral boson $\rho$.
Like the chiral boson $\phi$ in the RNS formalism, correlation
functions of $\rho$ can have unphysical poles which need to
be treated carefully. Fortunately, as in the special multiloop
amplitudes computed in \topamp\topo,
this subtlety can be ignored here because the chiral boson $\rho$
will decouple from the other worldsheet variables.

The topological prescription for the one-loop Type II superstring 
amplitude is
\eqn\presc{{\cal A}= \int d^2 \tau (\tau_2)^{-2}\int d^2 z_1 ...\int d^2 z_N
\langle (\int J_L\wedge J_R)^2 ~V_1(z_1,\bar z_1) ...
V_N(z_N,\bar z_N) \rangle}
where $\int J_L\wedge J_R = \int d^2 w (-\p \rho_L+ J_L^C)
(-\bar\p\rho_R + J_R^C)$ is constructed from the U(1) currents of
\hybfour\ and $\langle ~~\rangle$ signifies the 
two-dimensional correlation function on a torus. As discussed in
\topo, this prescription reproduces (up to picture-changing subtleties)
the standard one-loop RNS prescription in the large Hilbert space.
Note that when written in terms of RNS varaiables, $(\int J_L\wedge J_R)^2 = 
(\int d^2 w (b_L c_L +\xi_L\eta_L) (b_R c_R +\xi_R\eta_R))^2$, and
this term is necessary for providing the $[b_L,c_L,\xi_L,\eta_L]$
and $[b_R,c_R,\xi_R,\eta_R]$ zero modes in the large RNS
Hilbert space.

When the external states are d=4 supergravity states represented
by \closedv, the vertex operators are
independent of the $\rho$ variable and the compactification
variables. So the two-dimensional correlation function factorizes
into a d=4 contribution coming from the 
$[x^m,\t^\a,\t^\ad,p_\a,\bar p_\ad]$
worldsheet variables
which depends on the external states, and a d=6 contribution coming from the
$\rho$ variable and compactification variables which does
not depend of the external states.
Since the d=4 worldsheet variables are free fields, one can 
easily compute their correlation functions on a torus. Although
the compactification variables are not necessarily free fields, they
only contribute an overall factor through their partition
function which is independent of 
the external momenta and polarizations. 

\subsec{Partition functions on a torus}

Since the d=4 worldsheet variables
$[x^m,\t^\a,\bar\t^\ad,p_\a,\bar p_\ad]$ all are periodic free fields,
it is straightforward to compute their partition functions. From the 
four $x^m$'s,
the partition function is $(|\eta(\tau)|^{-2}\tau_2^{-\half})^4 =
|\eta(\tau)|^{-8}\tau_2^{-2}$ where $\eta(\tau) = e^{{\pi i\tau}\over{12}}
\prod_{n=1}^\infty(1-e^{2\pi i n\tau})$ is the Dedekind eta function.
And the partition functions for the 
fermions $[\t_L^\a,\tb_L^\ad,\t_R^\a,\tb_R^\ad]$ and $[p_{L\a},
\bar p_{L\ad},p_{R\a},\bar p_{R\ad}]$ contribute
$|\eta(\tau)|^{16}$ for correlation functions involving all sixteen
fermionic zero modes. Note that these sixteen fermionic zero modes
will come from the external vertex operators.

The partition functions of the remaining variables depend on the
twisted $\hat c=3$ N=2 superconformal field theory which describes
the compactification manifold. For example, for the uncompactified
superstring, the remaining variables are $[x^j,\bar x^{\bar j},$
$\Gamma_L^j, \bar \Gamma_L^{\bar j},$
$\Gamma_R^j, \bar \Gamma_R^{\bar j}, \rho_L,\rho_R]$ where $j=1$ to 3,
$\Gamma_{L/R}^j$ are fermions of zero conformal weight
and $\Gamma_{L/R}^{\bar j}$ are fermions of $+1$ conformal weight.

The partition function for the six uncompactified
$x$'s provides the factor
$(|\eta(\tau)|^{-2}\tau_2^{-\half})^6$
$ =$
$ |\eta(\tau)|^{-12}\tau_2^{-3}$. Naively, the partition function
for
$[\Gamma_L^j, \bar \Gamma_L^{\bar j},$
$\Gamma_R^j, \bar \Gamma_R^{\bar j}]$ is zero since 
$\langle (\int J_L\wedge J_R)^2\rangle =
\langle (\int (-\p\rho_L + \Gamma_L^j\bar\Gamma_L^{\bar j})
(-\p\rho_R + \Gamma_R^j\bar\Gamma_R^{\bar j}))^2\rangle$
provides at most eight of the twelve fermion zero
modes.
However, the partition function for the $\rho$ field diverges
in the absence of $e^\rho$ factors, so the expression 
$\langle (\int J_L\wedge J_R)^2\rangle $ needs to be regularized.
This can be easily done by writing
\eqn\onedef{1= \lim_{y\to z} \Gamma^1_L(y)\Gamma^1_R(\bar y)
e^{-\rho_L(y)-\rho_R(\bar y)}~~
\bar\Gamma^{\bar 1}_L(z)\bar\Gamma^{\bar 1}_R(\bar z)
e^{\rho_L(z)+\rho_R(\bar z)} }
and computing $\langle
~1 ~(\int J_L\wedge J_R)^2\rangle$. With this regularization,
the partition function for $[\Gamma_L^2,\Gamma_L^3,\bar\Gamma_L^{\bar 2},\bar
\Gamma_L^{\bar 3}]$
and $[\Gamma_R^2,\Gamma_R^3,\bar\Gamma_R^{\bar 2},\bar
\Gamma_R^{\bar 3}]$ gives $|\eta(\tau)|^8 \tau_2^2$ where the eight
fermion zero modes come from $(J_L\wedge J_R)^2$ and the $\tau_2^2$ factor
comes from integrating $(J_L\wedge J_R)$ twice over the torus.

Finally, the partition function for the 
$[\Gamma_L^1,\bar\Gamma_L^{\bar 1},\rho_L]$ and
$[\Gamma_R^1,\bar\Gamma_R^{\bar 1},\rho_R]$ variables can be
computed using the same method as was used in \ref\wata
{U. Carow-Watamura, Z.F. Ezawa, K. Harada, A. Tezuka and
S. Watamura, ``Chiral Bosonization of Superconformal Ghosts on
Riemann Surface and Path Integral Measure'', Phys. Lett. B227 (1989) 73.}
for the $[\xi_L,\eta_L,\phi_L]$ and
$[\xi_R,\eta_R,\phi_R]$ variables in the large RNS Hilbert space.
It was shown in \wata\ that functional integration over $[\xi,\eta,\phi]$
variables reproduces the Verlinde-Verlinde prescription of
\ref\vv{E. Verlinde and H. Verlinde, ``Multiloop Calculations in
Covariant Superstring Theory'', Phys. Lett. B192 (1987) 95.} for
$[\beta,\gamma]$ correlation functions if one inserts
$\xi_L(v)\xi_R(\bar v)(\int \eta_L \wedge \eta_R)^g$
into the $g$-loop correlation function. For example, the one-loop $[\b,\g]$
correlation function 
\eqn\bgone{\langle \d(\g_L(y))\d(\g_R(\bar y))~\d(\b_L(z))
\d(\b_R(\bar z))\rangle}
is reproduced by the $[\xi,\eta,\phi]$ correlation function
\eqn\xirep{
\langle \xi_L(v)\xi_R(\bar v)(\int d^2 w~ \eta_L(w)\eta_R(\bar w))
 e^{-\phi_L(y)-\phi_R(\bar y)} e^{\phi_L(z)+\phi_R(\bar z)}\rangle.}
Since only the zero modes of $\xi$ and $\eta$ contribute to \xirep,
this correlation function can be written as 
\eqn\xireptwo
{\tau_2\langle \xi_L(y)\xi_R(\bar y) \eta_L(z)\eta_R(\bar z)
e^{-\phi_L(y)-\phi_R(\bar y)} e^{\phi_L(z)+\phi_R(\bar z)}\rangle}
where the $\tau_2$ factor comes from the $d^2 w$ integration.
But 
\eqn\bgtwo{\langle
\d(\g_L(y))\d(\g_R(\bar y))~\d(\b_L(z))\d(\b_R(\bar z))\rangle 
=|\eta(\tau)|^{-4}}
for odd spin structure, which is the relevant spin
structure for the periodic $\rho$ variable. So by comparing with \xireptwo\
one finds that $\langle ~1~\rangle = |\eta(\tau)|^{-4} \tau_2^{-1}$ for the
$[\Gamma_L^1,\bar\Gamma_L^{\bar 1},\rho_L]$ and
$[\Gamma_R^1,\bar\Gamma_R^{\bar 1},\rho_R]$ partition functions where
$1$ is defined in \onedef.
  
Multiplying together the above partition functions, one finds
for the uncompactified superstring that all $\eta(\tau)$ factors
cancel and
\eqn\preamp{{\cal A}= \int d^2 \tau (\tau_2)^{-6}
\int d^2 z_1 ...\int d^2 z_N
\langle\langle  ~V_1(z_1,\bar z_1) ...
V_N(z_N,\bar z_N) \rangle\rangle}
where $\langle\langle~\rangle\rangle$ signifies the correlation function
on a torus divided by the partition function, and the correlation
function must include
all sixteen zero modes of the $d=4$ fermionic variables.
Using techniques similar to those used in \kn\ for tree amplitudes,
\preamp\ will now be explicitly computed.

\subsec{Koba-Nielsen-like formula for one-loop amplitude}

To evaluate \preamp\ for external massless compactification-independent
states, it is convenient to use \closedv\ to write
$$\langle \langle \prod_{r=1}^N V_r(z_r,\bar z_r)\rangle\rangle
=
\langle \langle \prod_{r=1}^N |H_r|^2 U_r(z_r,\bar z_r)\rangle\rangle$$
\eqn\firstv{=\prod_{s=1}^N {\p\over{\p\e_s}}
{\p\over{\p\bar\e_s}}
|_{\e_s=\bar\e_s=0} ~
\langle \langle | \exp (\sum_{r=1}^N \e_r H_r) |^2 \prod_{t=1}^N
U_t(z_t,\bar z_t)\rangle\rangle}
where $H_r$ is defined in \defHH\ 
and $D_{r\a}$ and $\bar D_{r\ad}$ are fermionic derivatives
which act only on $U_r(z_r,\bar z_r)$. 
For example,
$D_{2\a} \prod_{r=1}^N U_r(z_r,\bar z_r) $
$=
U_1(z_1,\bar z_1) ~D_\a U_2(z_2,\bar z_2)~\prod_{r=3}^N U_r(z_r,\bar z_r).$
Furthermore, it will be convenient to introduce the notation
\eqn\defffh{H_r=d_\a (z_r) w^\a_r 
+\bar d_\ad (z_r)\bar w^\ad_r
+\p\t_\a(z_r) a_r^\a
+\p\tb_\ad(z_r) \bar a_r^\ad +\Pi_m(z_r) b_r^m }
where
\eqn\defww{w_r^\a=-
(\bar D_r)^2 D_r^\a,~~
\bar w_r^\ad =-(D_r)^2\bar D_r^\ad,~~ 
a_r^\a= -D_r^\a,~~\bar a_r^\ad=
\bar D_r^\ad,~~ b_r^m={i\over 2}\s^m_{\a\ad} [D_r^\a,\bar D_r^\ad].}
Note that the $L/R$ index has been suppressed in these formulas
and that $|~~|^2$ signifies the left-right product.

The first step in evaluating \firstv\ is to eliminate the $d^\a$
variables using the OPE's of \dope.\foot{Although one could instead have
used the free field OPE's of \ope\ to eliminate the $p^\a$ variables
of \firstv, this would break manifest d=4 supersymmetry.} After
separating off the zero mode $d_{(0)}^\a$ of $d^\a$, any
correlation function involving $d^\a(z)$ is determined by the conditions
that it is a periodic function of $z$ and has the pole structure determined
by the OPE's of \dope.
So \firstv\ is equal to 
\eqn\sec{
\prod_{s=1}^N {\p\over{\p\e_s}} {\p\over{\p\bar\e_s}} |_{\e_s=\bar\e_s=0} 
\langle\langle:|\exp( M+ \sum_r \e_r d_{(0)\a} w_r^\a )|^2 : 
\prod_{r=1}^N U_r(z_r,\bar z_r)\rangle\rangle}
where
\eqn\defMM{
M=\sum_r \e_r[\bar d_\ad (z_r)\bar w^\ad_r
+\p\tb_\ad(z_r) \bar a_r^\ad +\Pi_m(z_r) b_r^m]}
$$+\sum_{r,s} \e_r w_r^\a 
F(z_s-z_r) (D_{s\a} + \e_s 
(i\Pi_{\a\ad}(z_s)\bar w^\ad_s -i\p\tb^\ad(z_s)
b_{s\a\ad}))$$
$$ - \sum_{r,s}\e_r\e_s \p F(z_r-z_s) w_r^\a a_{s\a}
+\sum_{r,s,t}\e_r\e_s\e_t F(z_r -z_s)F(z_t -z_s)~
\p\tb_\ad(z_s)\bar w^\ad_s w_{t\a}w^\a_r,
$$
the function $F(z)$ in \defMM\
is the Dirac
propagator for odd spin structure
$$F(z)=\p_z \log\Theta_1(z,\tau),$$
and the normal ordering symbol
$:~~:$ signifies that $D_r^\a$ is always
ordered to the left of 
$[w_r^\a,\bar w_r^\ad,a_r^\a,\bar a_r^\ad, b_r^m]$.
The term proportional to $\e_r\e_s\e_t$ in \defMM\
comes from the pole of $d^\a(z_t)$ with the residue of the pole
of $d^\b(z_r)$ at $z_s$. Note that 
all $\t^\a(z)$ 
variables in \sec\ can be set equal to their zero mode $\t^\a_{(0)}$ since
there are no more $d_\a$ variables to contract with.

Although $F(z)\to F(z)-2\pi in $ under $z\to z+m+n\tau$, the correlation
function of \sec\ is single-valued on the torus after integrating
out the fermionic zero modes $d_{(0)}^\a$. To see this, suppose that
$z_u\to z_u+m +n\tau$ for some $u$. Then 
$M\to M- 2\pi i n~\d M$ where
$$\d M= (\sum_r \e_r w_r^\a)
[D_{u\a} +\e_u
(i\Pi_{\a\ad}(z_u)\bar w^\ad_u -i\p\tb^\ad(z_u) b_{u\a\ad} +2
\p\tb_\ad(z_u)\bar w_u^\ad \sum_t \e_t w_{t\a} F(z_t-z_u))]$$
$$- \e_u w_u^\a 
\sum_s[D_{s\a} +\e_s (i\Pi_{\a\ad}(z_s)\bar w_s^\ad
-i\p\tb^\ad(z_s) b_{s\a\ad} +2
\p\tb_\ad(z_s)\bar w_s^\ad \sum_t \e_t w_{t\a} F(z_t-z_s))].$$
Since integrating over $d^\a_{(0)}$ brings down the term
$(\sum_r \e_r w_r^\a)^2$, the first term in $\d M$ does not contribute
since it is proportional to
$\sum_r \e_r w_r^\a$.
One can check that the second term in $\d M$ contributes
\eqn\seccon{\langle\langle :\d M |\exp M |^2: 
\prod_t U_t(z_t,\bar z_t)\rangle\rangle
=-\langle\langle :\e_u w_u^\a (\oint dy ~d_\a)|\exp M |^2 :
\prod_t U_t(z_t,\bar z_t)\rangle\rangle}
where the contour integration in $\oint dy ~d_\a$ goes around all
N external vertex operator locations $z_r$. Deforming this contour
off the back of the torus, one finds that \seccon\ is zero, so
the correlation function of \sec\ is single-valued on the torus.

One can similarly use the OPE's of \dope\ to eliminate
$\bar d^\ad$ in \sec\ and write 
$\langle \langle \prod_{r=1}^N V_r(z_r,\bar z_r)\rangle\rangle$ as
\eqn\third{
\prod_{s=1}^N {\p\over{\p\e_s}} {\p\over{\p\bar\e_s}} |_{\e_s=\bar\e_s=0} 
\langle\langle|:\exp( \sum_r \e_r (d_\a^{(0)} w_r^\a +
\bar d_\ad^{(0)} \bar w_r^\ad) + L):|^2  
\prod_{r=1}^N U_r(z_r,\bar z_r)\rangle\rangle}
where 
\eqn\defMprime{
L= 
\sum_r \e_r\Pi_m(z_r) b_r^m +
\sum_{r,s} \e_r F(z_r-z_s)
(D_{s\a}w_r^\a +\bar D_{s\ad}\bar w_r^\ad)}
$$+\sum_{rs} \e_r\e_s [F(z_r-z_s) 
i\Pi_{\a\ad}(z_s)\bar w^\ad_s w^\a_r + \p F(z_r-z_s)
(a_{s\a} w_r^\a
+\bar a_{s\ad}\bar w_r^\ad)]$$
$$-i
\sum_{r,s,t}\e_r\e_s\e_t 
b_{t\a\ad} w^\a_s \bar w_r^\ad F(z_t-z_s) \p F(z_r-z_t)
$$
$$+
\sum_{r,s,t,u}
\e_r\e_s\e_t\e_u \bar w_{t\ad} w_{u\a} w^\a_s \bar w_r^\ad
F(z_u-z_r) F(z_s-z_r) \p F(z_t-z_r),$$
all $\tb^\ad(z)$ variables are set equal to their zero mode $\tb^\ad_{(0)}$,
and the normal ordering symbol $:~~:$ now signifies that
$\bar D_r^\ad$ is always ordered to the left of
$D_r^\a$, which is always ordered to the left of 
$[w_r^\a,\bar w_r^\ad,a_r^\a,\bar a_r^\ad, b_r^m]$.

The final step in evaluating 
$\langle \langle \prod_{r=1}^N V_r(z_r,\bar z_r)\rangle\rangle$
is to perform the correlation function over the $x^m$ variables.
Note that $\Pi^m$ in \defMprime\ is equal to $\p x^m$
since $\t^\a$ and $\tb^\ad$ have been set equal to their zero modes.
The
correlation function over $x^m$ is easily performed using 
the scalar Green's function $x^m(y,\bar y) x^n(z,\bar z) \to \eta^{mn}
G(y-z)$ where
\eqn\greeng{G(y-z)= - \log |\Theta_1(y-z,\tau)|^2 + {{2\pi}\over{\tau_2}}
[Im (y-z)]^2.}
For example, 
\eqn\exgb{\langle\langle |\exp(\sum_r c_r^m \p x_m(z_r))|^2 \prod_s
e^{ik_s^m x_m(z_s)}\rangle\rangle }
$$= 
\exp(\sum_{r,s} [-\half k_r^m k_{sm} G(z_r-z_s) +
i c_r^m k_{sm}\p G(z_r-z_s) +i \bar c_r^m k_{sm}\bar \p G(z_r-z_s)$$
$$
-\half c_r^m c_{sm} \p^2 G(z_r-z_s)
-\half \bar c_r^m \bar c_{sm} \bar\p^2 G(z_r-z_s)
- c_r^m \bar c_{sm} \p\bar\p G(z_r-z_s)])$$
$$= \exp
(-{{2\pi}\over{\tau_2}} [Im \sum_r (c^m_r +i k_r^m z_r)][Im \sum_s
(c_{sm} +i k_{sm} z_s )])$$
$$
|\exp (-i c_r^m k_{sm} F(z_r-z_s)+\half c_{rm}c_s^m \p F(z_r-z_s)
+\half k_r^m k_{sm} \log\Theta_1(z_r-z_s,\tau))|^2$$
where $\bar k_r^m$ in \exgb\ is defined by $\bar k_r^m= - k_r^m$ so that
$\overline{e^{ik_r^m x_m(z_r)}} = 
e^{ik_r^m x_m(z_r)}$.

So after performing the correlation function over $x^m$, integrating out
the zero modes of $[d_{(0)}^\a,\bar d_{(0)}^\ad, \t_{(0)}^\a,\tb_{(0)}^\ad],$
and plugging into \preamp, one finally obtains the N=2 d=4
supersymmetric Koba-Nielsen-like formula for the one-loop amplitude
\eqn\ampf{{\cal A} = \int d^2\tau (\tau_2)^{-6}
\int d^2 z_1 ...\int d^2 z_N
\prod_{s=1}^N {\p\over{\p\e_s}} {\p\over{\p\bar\e_s}} |_{\e_s=\bar\e_s=0} }
$$\int d^2\t_L d^2\tb_L d^2\t_R d^2\tb_R ~: |\exp C|^2 :
(\sum_r\e_r w_{Lr})^2 (\sum_s\e_s \bar w_{Ls})^2 
(\sum_r\bar\e_r w_{Rr})^2 (\sum_s\bar\e_s \bar w_{Rs})^2 $$
$$\exp (-{{2\pi}\over{\tau_2}} [Im (K^m +i \sum_r k_r^m z_r)]^2 )
\prod_{r,s} |\Theta_1 (z_r-z_s,\tau)|^{k_r^m k_{s m}} \prod_t
\widetilde U_t(k,\t_L,\tb_L,\t_R,\tb_R)$$
where
\eqn\defMpp{C = 
 \sum_{r,s} \e_r F(z_r-z_s)
(D_{s\a}w_r^\a +\bar D_{s\ad}\bar w_r^\ad -i k_{s m} b_r^m) }
$$+\sum_{r,s,t} \e_r\e_s [F(z_r-z_s) F(z_s-z_t)
k_{t\a\ad}\bar w^\ad_s w^\a_r + \p F(z_r-z_s)
(a_{s\a} w_r^\a
+\bar a_{s\ad}\bar w_r^\ad +\half b_r^m b_{s m})]$$
$$-i
\sum_{r,s,t}\e_r\e_s\e_t 
b_{t\a\ad} w^\a_s \bar w_r^\ad (F(z_t-z_s)-F(z_r-z_s))
\p F(z_r-z_t) $$
$$+
\sum_{r,s,t,u}
\e_r\e_s\e_t\e_u \bar w_{t\ad} w_{u\a} w^\a_s \bar w_r^\ad
(F(z_u-z_r)- F(z_u-z_t)) F(z_s-z_r) \p F(z_r-z_t),$$
\eqn\kkdef{K^m = 
\sum_r \e_r b_r^m -i \s^m_{\a\ad}\sum_{r,s}
\e_r \e_s w_r^\a \bar w_s^\ad F(z_r-z_s),}
$\widetilde U_t(k,\t,\tb)$ 
is the Fourier transform of $U_t(x,\t,\tb)$,
$\bar k_r^m$ is defined by
$\bar k_r^m=- k_r^m$ in 
$\bar K^m$ and $\bar C$,
$F(z)=\p_z\log\Theta_1(z,\tau)$,
$[w_r^\a,\bar w_r^\ad,a_r^\a,\bar a_r^\ad, b_r^m]$ is defined in \defww,
and the normal ordering symbol $:~~:$ signifies that
$\bar D_r^\ad$ is always ordered to the left of
$D_r^\a$, which is always ordered to the left of 
$[w_r^\a,\bar w_r^\ad,a_r^\a,\bar a_r^\ad, b_r^m]$.

\newsec{Consistency of the One-Loop Formula}

Although the derivation of \ampf\ was completely straightforward,
the consistency of this one-loop formula will now be checked by
showing that it is invariant under gauge transformations of the external
states, single-valued as a function of the vertex operator locations,
modular invariant, and agrees with the RNS one-loop amplitude when
all external states are in the NS-NS sector.

\subsec{Gauge invariance}

To check the invariance of \ampf\ under the gauge transformation of
\ling, first consider $\d U_u = (D_L)^2 \L_{Lu}$ for the external state
at $z_u$. Since $D_L^\a \d U_u=0$, one sees from \defww\ that
$w_{Lu}^\a=\bar w_{Lu}^\ad=a_{Lu}^\a=0$, $\bar a_{Lu}^\ad= \bar D_{Lu}^\ad$,
and $b_u^m=-ik_u^m$. So from \defMpp\ and \kkdef, 
\eqn\gaugeder{
{\p\over{\p\e_u}} |_{\e_u=0} :|\exp C|^2 :
\exp (-{{2\pi}\over{\tau_2}} [Im (K^m +i \sum_r k_r^m z_r)]^2) }
$$
= :[-i\sum_r F(z_u-z_r) 
k_{rm}b_u^m +\sum_r
\e_r \p F(z_r-z_u)(\bar a_{u\ad}\bar w_r^\ad + b_r^m b_{um})$$
$$
-i\sum_{r,s} \e_r\e_s b_{u\a\ad} w_s^\a \bar w_r^\ad (F(z_u-z_s)-F(z_r-z_s))
\p F(z_r-z_u)$$
$$+{{ 2\pi i}\over{\tau_2}}b_{um} 
Im (K^m +i \sum_r k_r^m z_r)]
~|\exp C|^2
\exp (-{{2\pi}\over{\tau_2}} [Im (K^m +i \sum_r k_r^m z_r)]^2) :$$
$$=
:[-\sum_r F(z_u-z_r) 
k_{rm}k_u^m +\sum_r
\e_r \p F(z_r-z_u)(\bar D_{u\ad}\bar w_r^\ad -i b_r^m k_{um})$$
$$
+\sum_{r,s} \e_r \e_s k_{u\a\ad} w_s^\a \bar w_r^\ad F(z_r-z_s)
\p F(z_r-z_u)
+{{ 2\pi }\over{\tau_2}}k_{um} 
Im (K^m +i \sum_r k_r^m z_r)]$$
$$
|\exp C|^2
\exp (-{{2\pi}\over{\tau_2}} [Im (K^m +i \sum_r k_r^m z_r)]^2) :$$
where when changing $\bar a_{u\ad}$ to $\bar D_{u\ad}$, one needs
to be careful with normal ordering since 
\eqn\normalo{: e^{C}:\bar D_{u\ad} = 
: e^{C} \bar D_{u\ad}: - : [\bar D_{u\ad}, e^{C}] :}
$$=   
: e^{C} \bar D_{u\ad}: - :k_{u\a\ad}\sum_r\e_r F(z_r-z_u) w_r^\a
e^{C}: $$
when $\e_u=0$. One can easily check that \gaugeder\ is equal to 
\eqn\checkgauge{-{\p\over{\p z_u}}|_{\e_u=0}
:|\exp C|^2
\exp (-{{2\pi}\over{\tau_2}} [Im (K^m +i \sum_r k_r^m z_r)]^2) :
\prod_{r,s} |\Theta_1 (z_r-z_s,\tau)|^{k_r^m k_{s m}},}
so the amplitude of \ampf\ is gauge invariant up to a surface term.
Similarly, one can show that the gauge transformations
$\d U_u =  (\bar D_L)^2 \bar\L_L
+ (D_R)^2\L_R +(\bar D_R)^2 \bar\L_R$ only change \ampf\ by a surface
term.

\subsec{Single-valued function of vertex operator locations}

Under the transformation $z_u \to z_u + m + n\tau$ for integer $m$ and $n$,
the function 
$$F(z)=\p_z \log\Theta_1(z,\tau) \to F(z)-2\pi i n.$$
Since the points $z_u$ and $z_u+m+n\tau$ are identified on the torus,
one needs to check that the integrand of \ampf\ is single-valued
under this transformation. 

Using that $\sum_r \e_r w_r^\a =
\sum_r \e_r \bar w_r^\ad = \sum_r k_r^m=0$ from the
$[d_\a,\bar d_\ad, x^m]$ zero mode integration,
one finds that $:|\exp C|^2:$ transforms into 
\eqn\expmt{:|\exp(C + 2\pi i n[
\e_u (w_u^\a\sum_s D_{s\a} +\bar w_u^\ad \sum_s \bar D_{s\ad})}
$$
-i k_{um}\sum_r \e_r b_r^m +
\e_u w_u^\a \sum_{s,t}\e_s F(z_s-z_t)k_{t\a\ad} \bar w_s^\ad
+ k_{u\a\ad}\sum_{r,s}\e_r\e_s F(z_r-z_s)\bar w_s^\ad w_r^\a])|^2 :$$
$$=:|\exp(C + 2\pi  n k_{um} K^m)|^2 :$$
where the terms proportional to $\sum_s D_{s\a}$ and $\sum_s \bar D_{s\ad}$
are total derivatives which can be ignored after ordering $\sum_s D_{s\a}$
to the left of $e^{C}$ using 
\eqn\normaltwo{: e^{C} \sum_s D_{s\a}: = \sum_s D_{s\a} :e^{C}:
- :\sum_{r,s}\e_r F(z_r-z_s) k_{s\a\ad} \bar w_r^\ad  e^{C}:.}

Since the term 
\eqn\bost{\exp (-{{2\pi}\over{\tau_2}} [Im (i \sum_r k_r^m z_r)]^2) 
\prod_{r,s} |\Theta_1 (z_r-z_s,\tau)|^{k_r^m k_{s m}}}
is invariant under 
$z_u\to z_u 
+ m + n\tau$, 
$$\exp (-{{2\pi}\over{\tau_2}} [Im (K^m +i \sum_r k_r^m z_r)]^2) 
\prod_{r,s} |\Theta_1 (z_r-z_s,\tau)|^{k_r^m k_{s m}}$$
transforms into
$$\exp (-4\pi i n k_{um} Im K^m)
\exp (-{{2\pi}\over{\tau_2}} [Im (K^m +i \sum_r k_r^m z_r)]^2) 
\prod_{r,s} |\Theta_1 (z_r-z_s,\tau)|^{k_r^m k_{s m}}.$$
So the product 
$$:|\exp C|^2:
\exp (-{{2\pi}\over{\tau_2}} [Im (K^m +i \sum_r k_r^m z_r)]^2)
\prod_{r,s} |\Theta_1 (z_r-z_s,\tau)|^{k_r^m k_{s m}}$$
is invariant under $z_u\to z_u 
+ m + n\tau$, implying that the integrand of \ampf\ is single-valued.

\subsec{Modular invariance}

Under the modular transformation $\tau\to \tau' = -\tau^{-1}$
and $z\to z' = \tau^{-1} z$, the amplitude of \ampf\ should remain
invariant. To check this, it is useful to define $\e'_r =\tau^{-1}\e_r$. 
Using 
\eqn\suing{\tau'_2= |\tau|^{-2} \tau_2,\quad 
F(z,\tau)=\tau^{-1} F(z',\tau') -2\pi i z',\quad
\p F(z,\tau) = \tau^{-2} \p' F(z',\tau') -2\pi i\tau^{-1},}
one finds that
\eqn\finde{{\cal A}= \int d^2 \tau' (\tau'_2)^{-6} |\tau|^{-8}
\int d^2 z'_1 ...\int d^2 z'_N
\prod_{s=1}^N {\p\over{\p\e'_s}} {\p\over{\p\bar\e'_s}} |_{\e'_s=\bar\e'_s=0} }
$$\int d^2\t_L d^2\tb_L d^2\t_R d^2\tb_R : |\exp C|^2 :
(\tau\sum_r\e'_r w_{Lr})^2 (\tau\sum_s\e'_s \bar w_{Ls})^2 
(\bar\tau\sum_r\bar\e'_r w_{Rr})^2 (\bar\tau\sum_s\bar\e'_s \bar w_{Rs})^2 $$
$$\exp (-{{2\pi}\over{\tau_2}} [Im (K^m +i \sum_r k_r^m z_r)]^2 )
\prod_{r,s} |\Theta_1 (z_r-z_s,\tau)|^{k_r^m k_{s m}} \prod_t
\widetilde U_t(k,\t_L,\tb_L,\t_R,\tb_R)$$
where $C$ and $K^m$ are defined in \defMpp\ and \kkdef.

One can write $:\exp C:$ in terms of $\e'_r$, $z'_r$, and $\tau'$ using
\eqn\Mprimed{:\exp C(\e_r,z_r,\tau): =  
:\exp (C(\e'_r,z'_r,\tau') +2\pi i[ \sum_r \e_r z'_r (w_r^\a \sum_s D_{s\a}
+\bar w_r^\ad \sum_s \bar D_{s\ad})}
$$ -i\sum_r \e_r b_{r m} \sum_s k_s^m z'_s
-\sum_{r,s,t}\e_r \e_s (z'_r F(z_s-z_t) -z'_t F(z_r-z_s)) k_{t\a\ad}
\bar w_s^\ad w_r^\a$$
$$-\half\tau^{-1} \sum_{r,s}\e_r\e_s b_r^m b_{sm}
-i\tau^{-1}\sum_{r,s,t}\e_r\e_s\e_t
b_{t\a\ad} w_s^\a \bar w_r^\ad F(z_r-z_s)$$
$$-\tau^{-1} \sum_{r,s,t,u} \e_r\e_s\e_t\e_u \bar w_{t\ad} w_{u\a}
w_s^\a \bar w_r^\ad F(z_s-z_r)F(z_t-z_u)]):$$
$$=
:\exp (C(\e'_r,z'_r,\tau') +2\pi i\tau [-i K'_m\sum_t
k_t^m z'_t -\half K'_m K'^m]): $$
where 
\eqn\defkp{
K'^m = \tau^{-1} K^m = 
\sum_r \e'_r b_r^m -i \s^m_{\a\ad}\sum_{r,s}
\e'_r \e'_s w_r^\a \bar w_s^\ad F(z'_r-z'_s,\tau')}
and \normaltwo\ has been used to order $\sum_s D_{s\a}$ to the
left of $:\exp C:$. 

Since \bost\ is modular invariant, 
$$\exp (-{{2\pi}\over{\tau_2}} [Im (K^m +i \sum_r k_r^m z_r)]^2 )
\prod_{r,s} |\Theta_1 (z_r-z_s,\tau)|^{k_r^m k_{s m}}$$
$$= |\exp [{\pi\over 2}(\tau^2(\tau_2)^{-1}-(\tau'_2)^{-1}) 
(K'_m K'_m +2i K'_m\sum_s k_s^m z'_s)]|^2$$ 
$$\exp (-{{2\pi}\over{\tau'_2}} [Im (K'^m +i \sum_r k_r^m z'_r)]^2 )
\prod_{r,s} |\Theta_1 (z'_r-z'_s,\tau')|^{k_r^m k_{s m}}$$
$$=|\exp [\pi i\tau
(K'_m K'_m +2i K'_m\sum_s k_s^m z'_s)]|^2$$ 
$$\exp (-{{2\pi}\over{\tau'_2}} [Im (K'^m +i \sum_r k_r^m z'_r)]^2 )
\prod_{r,s} |\Theta_1 (z'_r-z'_s,\tau')|^{k_r^m k_{s m}}.$$
So 
$$:|\exp C(\e_r,z_r,\tau)|^2: 
\exp (-{{2\pi}\over{\tau_2}} [Im (K^m +i \sum_r k_r^m z_r)]^2 )
\prod_{r,s} |\Theta_1 (z_r-z_s,\tau)|^{k_r^m k_{s m}}$$
$$=
:|\exp C(\e'_r,z'_r,\tau')|^2: 
\exp (-{{2\pi}\over{\tau'_2}} [Im (K'^m +i \sum_r k_r^m z'_r)]^2 )
\prod_{r,s} |\Theta_1 (z'_r-z'_s,\tau')|^{k_r^m k_{s m}},$$
which implies using \finde\ that ${\cal A}$ is modular invariant.

\subsec{Equivalence with RNS amplitude for external NS-NS states}

In this final subsection, the one-loop amplitude of \ampf\ will
be shown to be equivalent to the RNS prescription when all
external d=4 states are in the NS-NS sector. The method used
for evaluating the RNS amplitude will be an N-point generalization
of the four-point one-loop computation in \ref\polch{J. Polchinski,
``String Theory, Volume 2: Superstring Theory and Beyond'',
Cambridge Univ. Press (1998).}. 

In the RNS formalism, the one-loop amplitude for $N$ external
massless d=4 NS-NS states can be written as
\eqn\arns{{\cal A}_{RNS}
= \int d^2\tau (\tau_2)^{-2} \prod_{r=1}^N\int d^2 z_r
\sum_{spin}\langle
(\int d^2 w ~b_L c_L(w)b_R c_R(\bar w)) 
\prod_{r=1}^N V^{RNS}_r(z_r,\bar z_r)\rangle}
where $V^{RNS}_r$ is defined in \vertrns, 
$\int d^2 w ~b_L c_L(w) b_R c_R(\bar w)$ provides the
fermionic ghost zero modes, and $\sum_{spin} \langle ~~\rangle$
signifies the two-dimensional correlation function summed over the
four possible spin structures on the torus. Since
$V_r$ only involves $\psi^m$ in four of the ten directions, the
odd spin structure does not contribute to \arns.

The first step to evaluating \arns\ is to compute the partition functions
of the various worldsheet fields. For the uncompactified superstring,
the $[\beta,\gamma]$ partition function
cancels the $[\psi^8,\psi^9]$ partition function, the $x^\mu$
partition function contributes $|\eta(\tau)|^{-20}\tau_2^{-5}$,
and the $[b_L,c_L,b_R,c_R]$ partition function contributes 
$|\eta(\tau)|^4 \tau_2$ where the $\tau_2$ factor 
comes from the $\int d^2 w$ integration in \arns.

To compute the partition function for $\psi^\mu$ for 
$\mu=0$ to 7, it is convenient to Wick-rotate to Euclidean space
and then use SO(8) triality to map $\psi^\mu$ to a Majorana-Weyl
SO(8) spinor variable $\chi^a$ for $a=1$ to 8. As in the GS formalism,
$\chi^a$ satisfies periodic boundary conditions on the torus so there is no
need to sum over spin structures. As discussed in \polch, the only subtlety
in this SO(8) triality map comes from the odd spin structure because of
the missing zero modes of $\psi^8$ and $\psi^9$. Note that the
odd spin structure
contributes parity-violating amplitudes involving odd powers of
the spacetime $\e$ tensor. However, as in
the one-loop four-point amplitude computed in \polch, there is no parity
violating contribution to one-loop N-point amplitudes when all external
states contain d=4 polarizations and momenta.
So the above subtlety coming from odd spin structures can be ignored.

Since the partition function for eight left and right-moving periodic
fermions contributes $|\eta(\tau)|^{16}$, the scattering amplitude of
\arns\ is 
\eqn\arnf{{\cal A}_{RNS}
= \int d^2\tau (\tau_2)^{-6} \prod_{r=1}^N\int d^2 z_r
\langle\langle
\prod_{r=1}^N V^{RNS}_r(z_r,\bar z_r)\rangle\rangle}
where 
\eqn\verr{V^{RNS}_r = 
(\p x^m +i\chi_L^a (\Gamma^{mp})_{ab}\chi_L^b k_p)
(\bar\p x^n +i\chi_R^c (\Gamma^{nq})_{cd}\chi_R^d k_q)
(h_{mn}+b_{mn}+\eta_{mn}\phi),}
$(\Gamma^{mn})_{ab}$ are constructed from the SO(8) Pauli matrices,
$\langle\langle ~~\rangle\rangle$
signifies the correlation function divided by the partition
function, and the correlation function must include all sixteen fermionic
zero modes of $\chi^a_L$ and $\chi^a_R$.

To compare \arnf\ with the one-loop amplitude of 
\preamp, divide the Majorana-Weyl SO(8) spinor $\chi^a$ into the
SO(4)$\times$ SO(4) spinors $\chi^{\a\b'}$ and $\chi^{\ad\bd'}$
for $(\a,\ad,\b',\bd')=1$ to 2 where the first $SO(4)$ acts on
the unprimed spinor indices and $\mu=0$ to 3, while the second SO(4) acts 
on the primed spinor indices and $\mu=4$ to 7.
Then by defining
\eqn\thdeff{\t^\a = \chi^{\a +'},\quad \tb^\ad =\chi^{\ad \dot -'},\quad
p^\a = \chi^{\a -'},\quad \bar p^\ad=\chi^{\ad \dot +'},}
one can equate the correlation function and zero modes of $\chi^a$
with the correlation functions and zero modes
of $[\t^\a,\tb^\ad,p_\a,\bar p_\ad]$.
Furthermore, one can use 
$$\chi^a (\Gamma^{mn})_{ab}\chi^b = p^\a (\s^{mn})_{\a\b} \t^\b +
\bar p^\ad (\bar \s^{mn})_{\ad\bd} \tb^\bd$$
to show that the
hybrid vertex operator of
\closedv\ for NS-NS states coincides with the RNS vertex operator of
\verr. So the RNS and hybrid one-loop amplitudes for external NS-NS
states have been proven to be equivalent.

\vskip 15pt
{\bf Acknowledgements:} NB would like to thank 
CNPq grant 300256/94-9,
Pronex grant 66.2002/1998-9,
and FAPESP grant 99/12763-0
for partial financial support.
BCV would like to thank FAPESP grant 00/02230-3 for financial support 
and the California Institute of Technology for hospitality.
This research was partially conducted during the period that NB was
employed by the Clay Mathematics Institute as a CMI Prize Fellow.

\listrefs
\end